\documentclass{Rinton-P9x6}

\def\bra#1{\langle #1 |}
\def\ket#1{| #1\rangle}
\def\braket#1#2{\langle \, #1 \, | \, #2 \, \rangle}

\def\R{\hbox{\rm I \kern-5pt R}}

\def\Tr{{\rm{Tr}}}
\begin{document}
\title{Large N Quantum Cryptography}
\author{Adrian Kent} 
\address{DAMTP, University of Cambridge, Centre for Mathematical
  Sciences,\\ Wilberforce Road, Cambridge CB3 0WA, U.K.\\
Email: a.p.a.kent@damtp.cam.ac.uk} 
\maketitle
\abstracts{In quantum cryptography, the level of security 
attainable by a protocol which implements a particular task $N$ times 
bears no simple relation to the level of security attainable by 
a protocol implementing the task once.   Useful partial security, 
and even near-perfect security in an appropriate sense, can be 
obtained for $N$ copies of a task which itself cannot be securely
implemented.  We illustrate this with protocols for quantum
bit string commitment and quantum random number generation between
mistrustful parties.} 
\section{Introduction} 
It is now well known that quantum information can guarantee 
classically unattainable security in a variety of important 
cryptographic tasks.  We know too from no-go results
that quantum cryptography cannot guarantee
perfect security for every task.  We cannot presently 
characterise precisely the tasks for which perfectly secure quantum 
protocols exist, or even the range of cryptographic 
tasks for which perfectly secure quantum protocols {\it might 
possibly} exist, because quantum cryptography involves
more than devising quantum protocols for tasks known to be useful
in classical cryptography.  The properties of quantum information 
allow new and cryptographically useful tasks, which
have no classical counterpart.  Also, reductions and relations
between classical cryptographic tasks need not necessarily apply 
to their quantum equivalents.  This means that there is
a wider range of tasks to consider, and that no-go theorems may 
not necessarily be quite as powerful as classical reasoning would suggest. 

These remarks apply in particular to bit commitment and coin tossing, 
important cryptographic protocols whose potential for physically secure
implementation has been extensively 
investigated. 
It is known that unconditionally secure quantum bit commitment is
impossible for non-relativistic 
protocols\cite{lochauprl,mayersprl,mayerstrouble,lochau,mayersone}:
that is, protocols in which the two parties are restricted to 
single pointlike sites, or more generally, in which the 
signalling constraints of special relativity are ignored.   
No unconditionally secure non-relativistic coin tossing protocol
has been found; no proof that no such protocols exist has yet been
published either.  

Unconditionally secure bit commitment is 
conjectured to be possible between parties controlling 
appropriately separated pairs of sites, when the impossibility of 
superluminal signalling is taken into
account.\cite{kentrel,kentrelfinite}
Unconditionally secure coin tossing is simple to implement under
these conditions.  However, we restrict attention to 
non-relativistic protocols in the rest of this paper, 
taking this as understood rather than inserting 
``non-relativistic'' throughout.  

Some variants of bit commitment, for which non-relativistic
protocols are not known to be impossible, have previously been  
studied.\cite{hk,atvy} 
We consider here a different generalisation, bit string 
commitment, in which one party commits many bits to another in a 
single protocol.  Two non-relativistic bit string commitment 
protocols, which offer classically
unattainable levels of security against cheating, are described.  

\section{Bit string commitment}  

Consider the following classical cryptographic problem. 
Two mistrustful parties, A and B, need a protocol which
will (i) allow A to commit a string $a_1 a_2 \ldots a_n$ of 
bits to B, and then, (ii) at any later time of her choice, 
reveal the committed bits.   The protocol should prevent $A$ 
from cheating, in the sense that she should have little or
no chance of unveiling bits $a'_i$ different from the 
$a_i$ without B being able to detect the attempted detection.
In other words, A should be genuinely committed after the
first stage.   The protocol should also prevent $B$ from
being able to completely determine the bit string.
More precisely, it must guarantee that, before revelation,
$B$ has little or no chance of obtaining more than $m$ bits of information 
about the committed string, for some fixed integer $m < n$.   

This {\it $(m,n)$ bit string commitment} problem
is a generalisation of the standard bit commitment problem, 
in which $n=1$ and $m=0$.  Clearly, a protocol for bit commitment
would solve this generalised problem, since the protocol could 
be repeated $n$ times to commit each of the $a_i$, and $B$ would
be able to obtain no information about the committed string.   
Conversely, classical reasoning implies that a protocol for the 
generalised problem, for any integers $m$ and $n$ with $m<n$, could be 
used as a protocol for standard bit commitment.  For $A$ and $B$ can use
any coding of a single bit $a$ by the $n$ bit string such that
none of the $m$ bits available to $B$ give information about $a$, 
and then use the protocol to commit $A$ to $a$.   

Classically, then, $(m,n)$ bit string commitment is essentially
equivalent to bit commitment.  However, there is no obvious 
equivalence between quantum $(m,n)$ bit string commitment and quantum bit 
commitment.  The impossibility
of unconditionally secure quantum bit commitment does not necessarily 
imply that, with an analogous definition of security, 
unconditionally secure quantum bit string commitment is impossible. 
In fact, the next sections show it can be achieved.  

\section{Protocol 1} 

Define qubit states $\psi_0 = \ket{0}$ and $\psi_1 = \sin \theta \ket{0} +
\cos \theta \ket{1}$, where $\sin^2 \theta = \delta$. 
We take $\theta > 0 $ and $r=n-m$ to be security parameters for the
protocol.   

{\bf Commitment:} \qquad 
To commit a string $a_1 \ldots a_n$ of bits to $B$, $A$ sends the 
qubits $\psi_{a_1} \, , \ldots \, , \, \psi_{a_n}$, sequentially.
\vskip 10pt
{\bf Unveiling:}\qquad 
To unveil, $A$ simply declares the values of the string bits, and
hence the qubits sent.  Assuming that $B$ has not disturbed the
qubits, he can test the bit values $a'_i$ claimed
by $A$ at unveiling by measuring the projection onto $\psi_{a'_i}$ 
on qubit $i$, for each $i$.  If he obtains eigenvalue $1$ in each
case, he accepts the unveiling as an honest revelation of a genuine
commitment.  If he obtains eigenvalue $0$ in any case, he concludes
(assuming that noise is negligible) that $A$ has cheated.  
\vskip 10pt
{\bf Security against A:} \qquad  Whatever strategy $A$ follows, 
once she transmits the qubits to $B$, their respective density
matrices $\rho_i$ are fixed.  Let $p^j_i = \bra{\psi_j} \rho_i \ket{\psi_j }$
be the probability of $B$ accepting a revelation of $j$ for the $i$-th
bit.  We have 
\begin{equation}
p^0_i + p^1_i \leq \cos^2 ( (\pi / 4 ) - ( \theta / 2)  ) + 
               \sin ( ( \pi / 4 ) + ( \theta / 2 ) ) \, , 
\end{equation}
which is $\leq 1 + \theta $ for small $\theta$.  
This is the standard definition of security against $A$ for an individual bit
commitment, with security parameter $\theta$.  
In other words, $A$'s scope for cheating on any bit of the string
is limited to slightly increasing the probability of revealing 
a $0$ or $1$, by an amount $\leq \theta$, which can be made arbitrarily
small by choosing the security parameters appropriately. 
\vskip 10pt
{\bf Security against B:} \qquad  
We assume that, prior to the commitment, $B$ has no information about 
the bit string and regards every possible value as equiprobable.   
From $B$'s perspective, then, he has to obtain information about
a density matrix of the form 
\begin{equation}
\rho = ( 1 / 2^n ) \sum_{a_1 \ldots a_n} \ket{ \psi_{a_1} \ldots \psi_{a_n} }
\bra{ \psi_{a_1} \ldots \psi_{a_n} } \, .
\end{equation}
Holevo's theorem\cite{holevo} 
tells us that the accessible information available to
$B$ by any measurement on $\rho$ is bounded by the entropy 
\begin{eqnarray}
\lefteqn{ S ( \rho )= 
( ( ( 1 + \sin \theta ) / 2 ) \log_2 ( ( 1 + \sin \theta ) / 2 ) + 
} \nonumber \\
&&  ( ( 1 - \sin \theta ) / 2 ) \log_2 ( ( 1 - \sin \theta ) / 2 ) )^n \, . 
\end{eqnarray}
Now, for any fixed $\theta > 0 $, we have $S(\rho ) < n $.  For any
fixed $r$, by taking $n$ sufficiently large, we can ensure 
$ n - S( \rho ) > r$.   In other words we can ensure that, 
however $B$ proceeds, an average of at least $r$ bits of information
about the string will remain inaccessible to him.  By choosing $n$
suitably large, we can also ensure that the probability of his
obtaining more than $n-r$ bits of information about the string is 
smaller than $\epsilon$, for any given $\epsilon>0$.   

A more efficient version of this protocol can be devised using 
qutrit states\cite{sr} --- an observation I owe to Rob Spekkens. 

\section{Protocol 2} 

Protocol 1 ensures bit-wise security against $A$, but uses 
a rather inefficient bit string coding which allows $B$ to obtain 
almost all of the bit string before revelation.     
For large $n$, more efficient codings allow the security against $B$ 
to be greatly enhanced, though with a weakened notion of security
against $A$.   

We again take $\theta > 0$ to be a security parameter and write
$\epsilon = \sin \theta$.  Now, for any $\theta > 0$ and large $n$,
explicit constructions are known for sets $v_1 , \ldots, v_{f(n)}$ of
vectors in $H^n$ such that $ | \braket{ v_i}{ v_j } | < \sin \theta $
for all $i \neq j$, with the property that $f(n) = O ( \exp ( C n )
)$, where $C$ is a positive constant that depends on
$\theta$.\cite{conwaysloane,justesen} (The use of these constructions
for efficient quantum coding of classical information has previously
been noted by Buhrman et al.\cite{bcww}, who describe efficient
quantum fingerprinting schemes which reduce communication complexity
in the simultaneous message passing model.)  A string of $O( C n )$
bits can thus be encoded by vectors in $H^n$, such that the overlap
between the code vectors for two distinct strings is always less than
$\sin \theta$, suggesting the following bit string commitment
protocol.

{\bf Commitment:} \qquad 
Let $N$ be the number of bits that can be encoded in $H^n$ by the
above construction.  To commit a 
string $a_1 \ldots a_N$ of bits to $B$, $A$ sends the 
state $v_{a_1 \ldots a_N}$, treating the index as a binary number.
\vskip 10pt
{\bf Unveiling:}\qquad 
To unveil, $A$ simply declares the values of the string bits, and
hence the state sent.  Assuming that $B$ has not disturbed the
qubits, he can test $A$'s claim at unveiling by measuring the projection
onto $v_{a_1 \ldots a_N}$.  If he obtains eigenvalue $1$, he accepts 
the unveiling as an honest revelation of a genuine
commitment.  If he obtains eigenvalue $0$, he concludes
that $A$ has cheated.  
\vskip 10pt
{\bf Security against A:} \qquad  As before, 
once $A$ transmits a quantum state to $B$, its density
matrix $\rho$ is fixed.  
Consider some set $i_1 , \ldots , i_r$ of bit strings which 
$A$ might wish to maintain the option of revealing
after commitment.  Let $P_i$ be the projection onto $v_i$, let $p_i = 
\Tr (\rho P_i )$ be the probability of $A$ successfully revealing 
string $i$, 
and write 
\begin{equation} Q = P_{i_1} + \ldots + P_{i_r} \, . \end{equation}
It is not too hard  to verify that 
\begin{equation}
\Tr ( \rho Q ) \leq 1 + (r-1) \epsilon 
\end{equation}
In other words, 
\begin{equation}
p_{i_1} + \ldots + p_{i_r} \leq 1 + f ( \epsilon , r ) \, , 
\end{equation}
where, for any fixed $r$, $f$ can be made as small as desired by
choosing $\theta$ suitably small.  

So, given that $A$ is determined to reveal a bit string from
some finite set of size $r$, her scope for cheating is limited to increasing
the probability of revealing any given element of the set
by a fixed amount.  For any fixed $r$, that amount can be made 
arbitrarily small by choosing the security parameters appropriately. 
If $B$'s concern is to prevent cheating of this type, for some
predetermined $r$, the protocol can guarantee him security. 
\vskip 10pt
{\bf Security against B:} \qquad  Holevo's theorem implies
that the information about the $N \approx C n$ bit string accessible to
$B$ is at most $\log n$ bits.   

\section{Asymptotically secure coin tossing}

Consider the following non-relativistic protocol for generating a string
of $N$ random bits between mistrustful parties.  We assume that 
$N$ is large, and take $M$ also to be large, with $\log M \ll N$.  
A prepares $M$ batches of $N$ Bell singlet states, and sends one particle
from each of the $MN$ singlets to B.   
B chooses $(M-1)$ of the batches, and asks A to send the second 
particle from each of the $(M-1)N$ singlets in these batches.
B tests that these $(M-1)N$ pairs of particles are indeed singlets.
If not, he concludes that A is cheating, and the protocol ends. 
If so, he accepts that A is honest.   A and B then use the last
batch of singlets to generate $N$ random bits, by carrying out 
correlated measurements (say of $\sigma_z$) and converting the
results to a bit string using a previously agreed protocol.  

{\bf Security against A:} \qquad  A can only cheat by preparing 
non-singlet states which bias the outcomes towards those she
would prefer.  Her scope for cheating is limited by the 
cut-and-choose step of the protocol, which ensures that, if 
any batch has low fidelity to $N$ singlet states, her
cheating will almost surely be detected. 

{\bf Security against B:} \qquad  B can cheat by carrying out
measurements on every particle from every batch sent to him,
deciding which batch gives the bit string most favourable for
his purposes, and choosing the other $(M-1)$ for the test.
However, this will allow him to fix only $\approx \log M$ bits
of information about the $N$ bit string.  With suitable $M,N$
this is an insignificant fraction.

\section*{Acknowledgments}
Much of this work was carried out while a visiting academic researcher
in the QIP Group at Hewlett-Packard Labs, Bristol.  I thank
Serge Massar, Bill Munro, Tim Spiller, Rob Spekkens, Terry Rudolph 
and Alain Tapp for helpful discussions.
This work was partially supported by the European collaboration EQUIP.

\end{document}